\documentclass[aps,prl,reprint]{revtex4-1}
\usepackage{blindtext}
\usepackage[colorlinks=true, allcolors=blue]{hyperref}
\usepackage{graphicx}
\usepackage{dcolumn}
\usepackage{siunitx} 
\newcommand{\rom}[1]{\uppercase\expandafter{\romannumeral #1\relax}}

\def\HL#1{#1}
\def\RM#1{#1}
\def\AD#1{#1}

\usepackage[utf8]{inputenc}
\usepackage[T1]{fontenc}
\usepackage{mathptmx}
\begin{document}
\title{Single Phonon Detection for Dark Matter via Quantum Evaporation and Sensing of $^3$Helium}

\author{S. A. Lyon, Kyle Castoria}
\author{Ethan Kleinbaum}
\altaffiliation{Current address: Honeywell Corp. Minneapolis, MN 55422}
\affiliation{Department of Electrical and Computer Engineering, Princeton University,
Princeton, NJ 08544}
\author{Zhihao Qin, Arun Persaud, Thomas Schenkel}
\affiliation{Acceleration Technology \& Applied Physics, Lawrence Berkeley National
Laboratory, 1 Cyclotron Road, CA 94720, USA}
\author{Kathryn M. Zurek}
\affiliation{Walter Burke Institute for Theoretical Physics, California Institute of
Technology, Pasadena, CA 91125}

\begin{abstract}
Dark matter is five times more abundant than ordinary visible matter in our
Universe.  While laboratory searches hunting for dark matter have traditionally
focused on the electroweak scale, theories of low mass hidden sectors motivate
new detection techniques.  Extending these searches to lower mass ranges,
well below $\SI{1}{\GeV\per c^2}$, poses new challenges as rare
interactions with standard model matter transfer progressively less energy
to electrons and nuclei in detectors.  Here, we propose an approach based
on phonon-assisted quantum evaporation combined with quantum sensors
for detection of desorption events via tracking of spin coherence.  The
intent of our proposed dark matter sensors is to extend the parameter
space to energy transfers in rare interactions to as low as a few meV
for detection of dark matter particles in the keV/c$^2$ mass range.  

\end{abstract}

\maketitle

Dark matter (DM) direct detection experiments have focused on detecting
Weakly Interacting Massive Particles (WIMPs) via nuclear recoils
(see {\em e.g.} Ref.~\cite{Cushman:2013zza} for a review), where DM
with mass in the 100 GeV range deposits energy by elastic
scattering.  However, in theories with low-mass hidden sectors (called
a hidden valley), thermal DM can be much lighter, even down to a keV
in mass where it carries meV of kinetic energy ($\frac{1}{2} m_X v_X^2$,
with $v_X \simeq 10^{-3} c$).  As the mass of the DM drops below
approximately 10 GeV, the detection of rare scattering events with
target nuclei falls below detection thresholds, and target nuclei
absorb a very small fraction of the DM kinetic energy; see
Ref.~\cite{Battaglieri:2017aum} for a review. At lower energies,
electron recoils with energy transfer thresholds in the $\SI{1}{\eV}$ range
can be detected with sensitive charge coupled devices (CCD) counting
electron-hole pairs in semiconductors, ({\em e.g.}~\cite{PhysRevLett.125.171802})
or athermal phonon detectors ({\em e.g.}~\cite{SuperCDMS:2020aus}). However,
dark matter events have not yet been observed in these energy ranges,
and it is desirable to probe thermal DM as light as 1 keV.  Thus developing
systems which can detect rare events with even lower deposited energy
is an important  goal.

In solids and liquids the lower energy excitations are generally
phonons~\cite{Knapen:2017ekk} (and rotons in superfluid
helium~\cite{Schutz:2016tid,PhysRevD.100.092007}).
Ionic crystals
(polar materials) are especially interesting as detectors, since they
enable new pathways for interaction with
DM~\cite{Knapen:2017ekk, Trickle:2019nya,PhysRevD.101.055004, PhysRevD.102.095005}. 
One
challenge to sensing these phonons is that they are itinerant.
Initially generated
optical phonons rapidly decay to acoustic phonons, which disperse
the deposited energy throughout the detection medium.
The development
of very sensitive and optimized detectors for
quasiparticles and phonons
using transition edge sensors (TES) and superconducting
nanowire detectors (SNSPD) is underway~\cite{Hochberg:2015fth}.

\begin{figure}[t]
	\includegraphics[width=\linewidth]{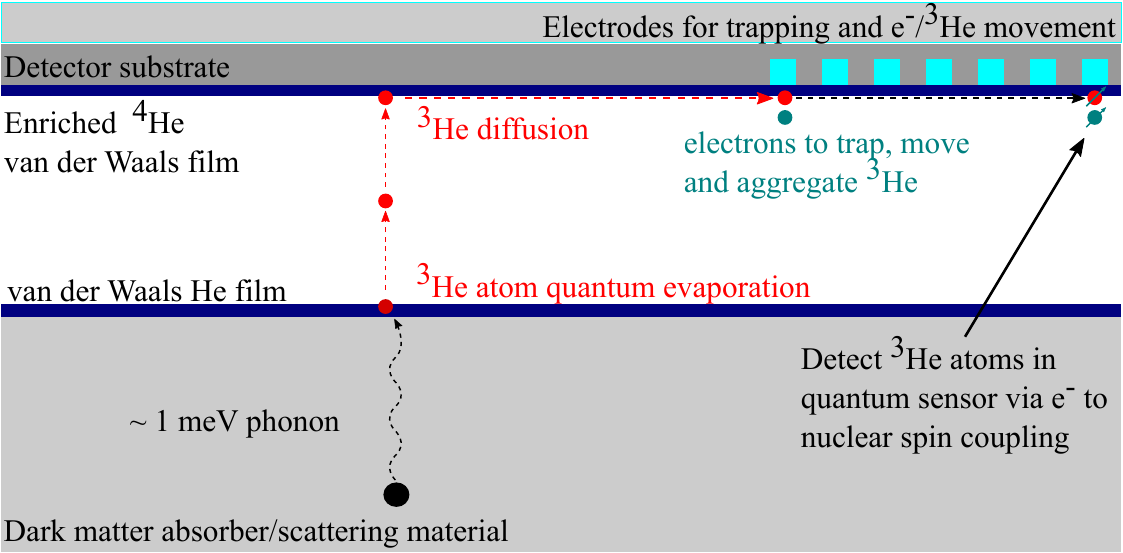}
	\caption{Schematic of the DM detector concept. An interaction with
	DM in an ionic crystal generates $\sim$1 meV phonons, which impinge
	on a surface covered with a van der Waals helium film. The phonon
	quantum evaporates a $^3$He atom from the surface of the film,
	which is then collected on the van der Waals film covering the
	detector structures. The $^3$He atoms diffuse until captured by
	an electron bound to the helium surface in a CCD-like structure.
	Periodically the collected $^3$He atoms are moved with the CCD
	to a readout device which operates via nuclear spin induced decoherence
	of an electron in a spin based quantum sensor.}
	\label{fig:detector_schematic}
\end{figure}

\HL
{
Here we propose an alternative, novel detection
concept for single low-energy phonons based on the quantum sensing
of the spin of $^3$He atoms which have been 
evaporated from the surface of a He van der Waals film coating
and ionic crystal.
}
This is related to
earlier proposals based upon He quantum
evaporation~\cite{PhysRevLett.119.181303, PhysRevD.100.092007},
though here we consider $^3$He which is bound to a $^4$He surface with
an energy of $\sim$5 K~\cite{Edwards1978}, somewhat less than the $\sim$7 K
binding of a $^4$He atom. More importantly, the nuclear spin of $^3$He
allows its quantum
sensing at the level of single atoms.

A diagram of this concept is shown in Fig.~\ref{fig:detector_schematic}.
There are four major steps in the dark matter detector proposed
here and shown in the Figure: (1) production of phonons through
the interaction with dark matter leading to the quantum evaporation
of $^3$He atoms from Andreev bound states~\cite{osti_4525526};
(2) trapping the $^3$He on the detector
surface using electrons bound to a film of isotopically enriched
liquid $^4$He; (3) collecting and transporting the electrons and trapped
$^3$He atoms to a detector structure; and (4) quantum sensing of the
$^3$He atoms through their nuclear spin. An important feature of this
detection concept is the separation of the dark matter absorber
({\em i.e.} target, such as a polar material) and the $^3$He detector,
which opens the possibility to readily select and test a series of
absorber materials for specific dark matter searches. Further, this
approach is compatible with large magnetic fields, a feature that
again enables testing of specific modes of proposed dark matter
interactions.  In addition, the disk-like form factor of the
absorber-sensor package that we envision as shown schematically
in Fig.~\ref{fig:detector_schematic} might enable future adaptation of our
concept using similar device integration concepts as developed {\em e.g.} in SuperCDMS. 
In the remainder of this paper, we describe each of the steps of our
detector concept in detail.
\HL
{
It spans a range of fields, including dark
matter astrophysics, solid state physics for phonon propagation, quantum fluids
for $^3$He evaporation, device physics for $^3$He trapping and transport,
and quantum information
and sensing for $^3$He detection.
}
Here we show how it can be a viable complement
to existing efforts for light DM detection with TESs, SNSPDs, and CCDs.

\paragraph{Helium evaporation via DM-produced phonons}---\,
\AD
{
{\em Bulk} superfluid He has been proposed for DM detection through the
production of phonons and rotons~\cite{PhysRevLett.119.181303, PhysRevD.100.092007}.
By contrast, here we propose to use the helium as a means to detect the phonons
produced in a solid target, and not as the target itself. This approach was  also discussed
in Ref.~\cite{PhysRevLett.119.181303}, though here we are specifically
suggesting {\em polar} targets, such as NaI. Except
for evidence that rotons do not efficiently evaporate $^3$He~\cite{Warren2000},
the remainder of this approach to detecting low-energy DM interactions
could be utilized for bulk He. However, there are important advantages
and complementary opportunities to interacting and generating phonons in
crystals (notably reach to a broader range of dark matter theories and masses~\cite{Trickle:2019nya,PhysRevD.101.055004}), when evaporating $^3$He from these.
}
We will focus on the case that a DM particle produces a single high-energy
($\gtrsim$ 10 meV)
phonon by an interaction with an ion in a polar material target.
The anticipated DM interaction rate is about 2/min. in a
1 kg NaI crystal (detailed theoretical
calculations can be found in Refs.~\cite{PhysRevD.101.055004,Trickle:2019nya}), with an expected  background from radioactive and
cosmogenic species about 50 times lower.
The appendices, include more detailed
discussions of (1) detector crystal criteria, interaction rates, and backgrounds
(2) other $^3$He detection approaches, and (3) possible alternative
adsorbates.

Below about $\SI{80}{\milli\kelvin}$ a He surface is covered with
$^3$He, both for bulk He and a van der Waals film. The athermal acoustic phonons
resulting from the decay of the high-energy phonon,
when interacting with the surface of the polar crystal coated with a
thin helium film, can lead to quantum evaporation.
Heat pulse experiments with natural abundance He films on
crystalline substrates
have shown that about $5\%$ of the detected atoms are directly
evaporated by phonons from the heat pulse -- the ''phonoatomic'' effect
depicted in Fig.~\ref{fig:detector_schematic}, while the remainder are
evaporated by the overall temperature rise of the
crystal~\cite{PhysRevLett.54.2034,More1996}.
However, these experiments have mostly used polished, rather than vacuum-cleaved surfaces.
It is known that even well-polished surfaces covered with helium lead to enhanced
phonon thermalization~\cite{phononsurfacethermalization} and inefficient transport of
phonons across the interface into a film~\cite{PhysRevLett.24.1049}. The efficiency
of quantum evaporation from a van der Waals film of liquid helium on a freshly
cleaved surface which has been protected from oxygen and humidity is not known. Boosting the
evaporation efficiency may also be possible by depositing a thin film of Cs on a crystal
and coating that with a monolayer of $^3$He, as suggested in Ref.~\cite{More1996}, since
$^3$He is bound to Cs by only about 2.4 K.

\paragraph{$^3$Helium trapping}---\,
As shown in Fig.~\ref{fig:detector_schematic}, the evaporated $^3$He atoms will
be collected on an adjacent helium-covered surface.  The helium in this
collector film will be isotopically enriched to remove its $^3$He. Enrichment
of $^4$He to less than 5 parts in $10^{13}$ (< 0.5 ppt) $^3$He has been
demonstrated~\cite{HENDRY1987131}.
\AD
{
The
enriched $^4$He film on this collector structure must be fully isolated from
the $^3$He/$^4$He mixture coating the DM target crystal.  There are two
well-established approaches to breaking a van der Waals film: a film-burner
as was employed in the HERON experiment~\cite{HeronFilmBurner1992};
and a band of cold-evaporated
Cs, since superfluid $^4$He does not wet Cs~\cite{Nacher1991, Taborek1992}.
Here we expect that the Cs film will be preferable, since the film-burner could
preferentially evaporate $^3$He atoms, which would appear as false events.
}
After being captured onto this enriched
$^4$He film, the $^3$He atoms diffuse across the film surface~\cite{PhysRevLett.77.2973}.
Our concept uses electrons held a few nanometers above the surface of the
helium film by applied electric fields to localize $^3$He atoms in dimples
under the electrons~\cite{WILLIAMS197135} and enable
their transport to spin readout sensors for detection.  It is essential
that the $^3$He atoms be localized
in the dimples for spin based $^3$He sensing, since if the $^3$He
atoms are allowed to diffuse freely, motional
narrowing causes
them to have little effect on an electron's
spin in a quantum sensor~\cite{PhysRevA.74.052338}.

Trapping $^3$He in dimples under electrons bound to superfluid $^4$He is newly suggested here, and arises from $^3$He reducing 
the surface tension at mK temperatures~\cite{osti_4525526}.
The addition of a $^3$He atom will deepen the dimple, lowering
the electron in the applied electric field, increasing its potential energy and
 trapping the $^3$He.  The depth and shape of the dimple in the
He surface will be determined by the
equilibrium between electrostatic forces pulling the electron against the
helium and capillary forces resisting the deformation.

\begin{figure}[t]
	\includegraphics[width=\linewidth]{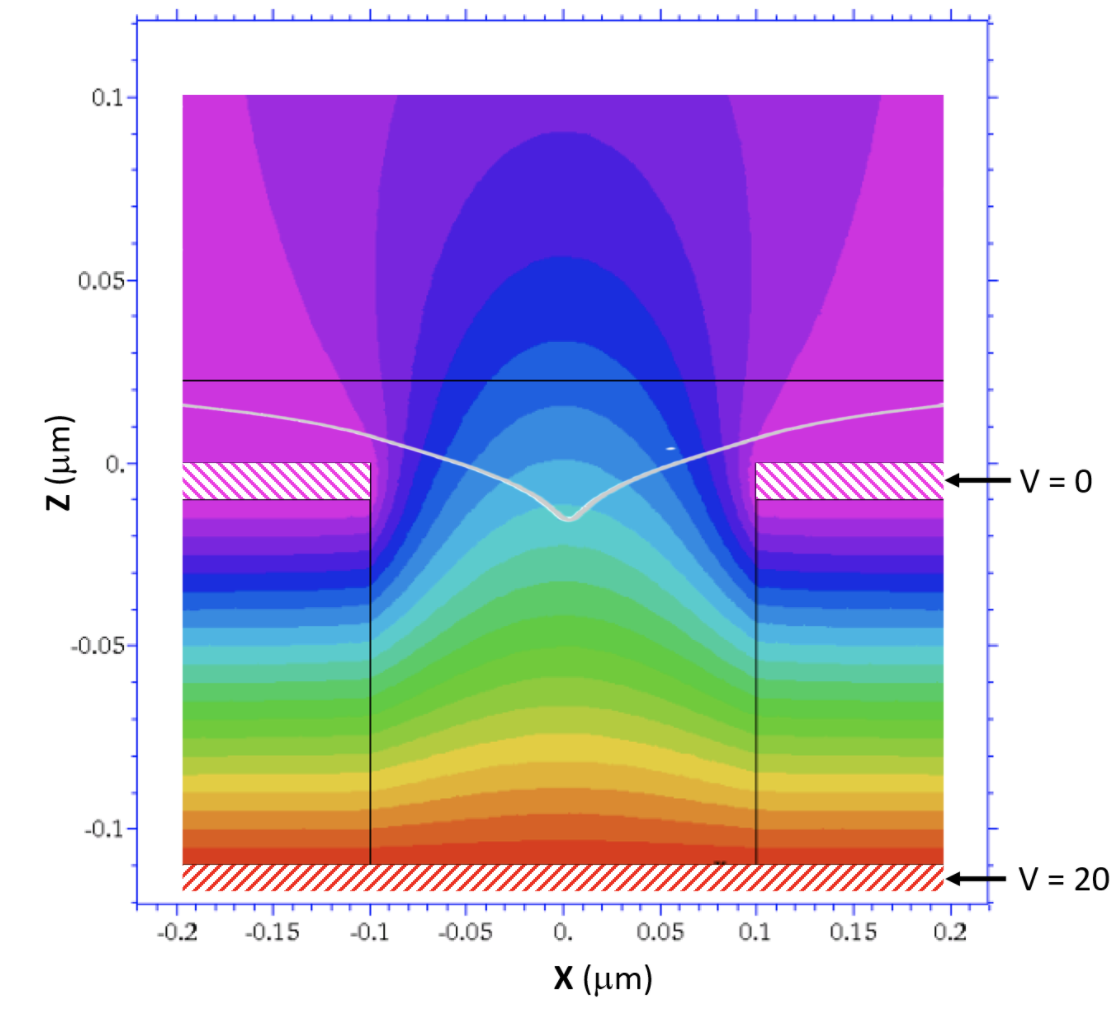}
	\caption{Finite element calculation of the potential in a 0.2 $u$m wide
	He-filled channel with metal gates biased as shown. The channel is assumed
	to extend in the Y-direction (into the page) and electrons are placed with
	a periodicity of 0.2 $u$m. The contours are at 1 V steps. The black horizontal
	line at a height of $\sim$0.02 $u$m shows the helium surface without the electron or electric fields.
	The white curve superimposed
	on the potential image is the calculated helium surface for an electron held
	in the channel with the applied voltages.}
	\label{fig:potential}
\end{figure}

To understand the temperature needed for stable trapping of the $^3$He
we have performed numerical calculations
of the helium dimple as shown in Fig.~\ref{fig:potential}. 
These devices will use ''channel'' technology~\cite{Marty1986StabilityOT, VANHAREN1998656}
in which an underlying metal layer is first deposited on a substrate and
patterned to make gate electrodes, and this layer of electrodes is then
covered with an insulator and a second metallic layer. This upper metal
layer is patterned lithographically, and areas are removed to form the
channels where the electrons will reside. With a small amount of bulk superfluid,
the helium covers the device through capillary action and fills the channels. Electrons emitted from the vacuum
with a positive bias on the underlying gate electrodes accumulate on
the helium film in the channels.

 Fig.~\ref{fig:potential} shows calculated electrostatic potentials for a
 $\SI{200}{\nm}$ wide channel that is $\SI{110}{\nm}$ deep. The lower metal
 electrode is biased to $+\SI{20}{\V}$, and the upper metal is at ground
 ($\SI{0}{\V}$), with the potential contours at $\SI{1}{\V}$ steps.
 The
 change in the dimple with the addition of a $^3$He atom is too small to be
 seen in the figure, but the vertical electric field is calculated to be about
 $\SI{0.8e6}{\V\per\cm}$ at the electron, so a very small change in dimple
 depth can produce a significant change in electrostatic energy. For the
 parameters of Fig.~\ref{fig:potential}, the calculated energy change per
 $^3$He atom is about $\SI{27}{\K}$.
 \AD
 {
 A variety of channel geometries and applied
 voltages have been modeled: higher voltages are required for narrow channels
 where capillary forces are stronger, while the helium surface becomes unstable if
 the channel becomes too wide. The calculations suggest that stable trapping of
 $^3$He is possible over at least a factor of 4 range in channel widths.
 }

It is expected that this detector will be operated at $\sim\SI{35}{\milli\kelvin}$,
or colder, since the background pressure of $^3$He must be kept very low. At this
temperature, if the trapping energy is \SI{2}{\K}, the calculated
density of free $^3$He atoms is ~$\sim\SI{e-12}{\per\cm\squared}$ for
every trapped $^3$He atom. Thus, trapping energies in
the range of \SIrange[range-phrase = --,range-units = single]{1}{2}{\K} will be
sufficient for localizing the $^3$He atoms.

\begin{figure}[t]
	\includegraphics[width=\linewidth]{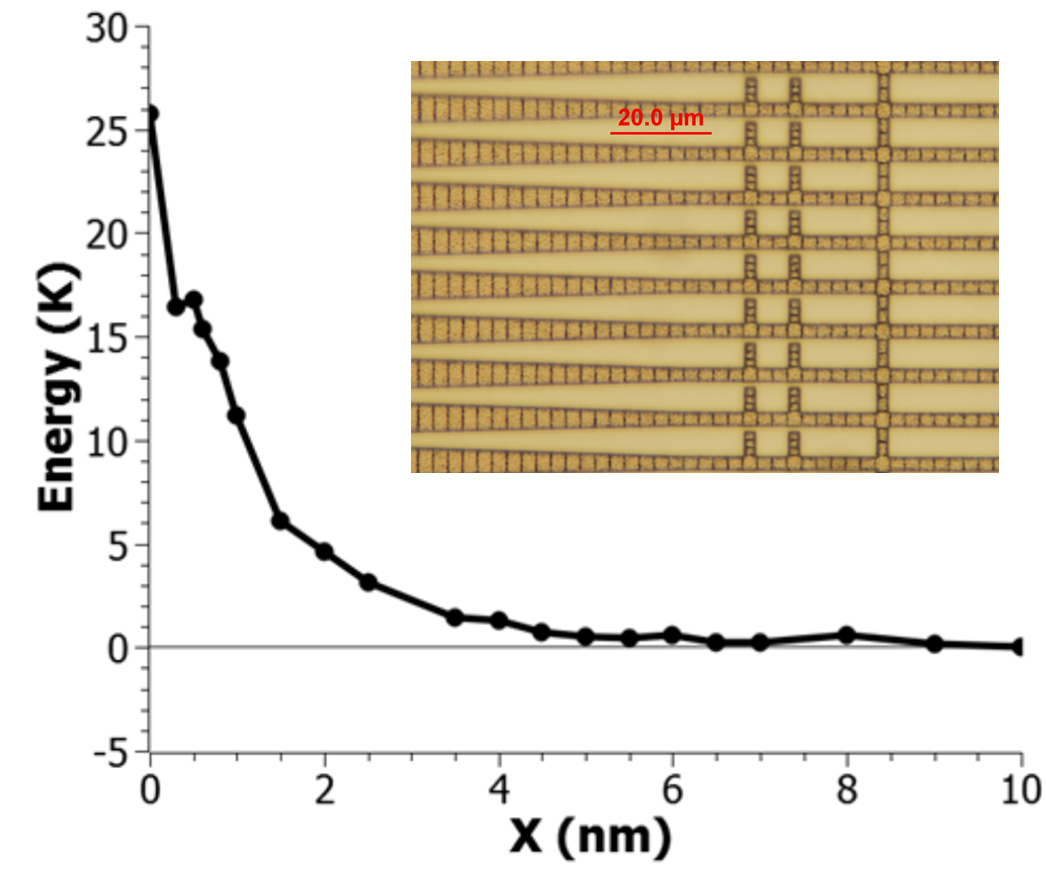}
	\caption{Calculation of the binding energy of a $^3$He atom to an electron
	in a channel like that shown in Fig. 3, but with the $^3$He displaced by a
	distance (X) in the X-direction (across the channel). Lines connecting dots are guides to the eye. The inset shows a CCD used for moving electrons along
	helium filled channels (the micrograph shows the metal layers). The main
	channels run horizontally and
	are  similar, though wider and deeper, to the CCDs needed for the detectors.
	The underlying gate electrodes
	run vertically as seen at the bottoms of the channels.
	(After Ref. \cite{PhysRevLett.107.266803}) }
	\label{fig:energy-ccd}
\end{figure}

Another consideration is the cross section for an electron to capture a $^3$He
atom. Again, we have calculated the cross section numerically, here by
introducing a change in the surface tension some distance from the
electron to determine how the energy changes. Results for these same
parameters (\SI{0.2}{\micro\meter} channel with a $\SI{20}{\V}$ bias)
are shown in Fig.~\ref{fig:energy-ccd}.  Taking the
criterion for capture that
the energy falls below kT, then we have a capture distance of $\sim\SI{6}{\nm}$.
In the orthogonal (Y) direction, calculations show that the capture
distance is smaller, of order $\SI{1}{\nm}$. This can be readily understood,
since the surface curvature is large in the X-direction, where van der Waals
forces require the helium surface to bend tightly around the edges of the
channel, as seen in Fig.~\ref{fig:potential}. In the Y-direction the channel
is long, the dimple is more gradual, and the change in surface tension is
only felt very close to the electron.

The $^3$He forms a Fermi gas on the $^4$He surface at low densities, and
its motion is diffusive. Measurements of the spin diffusion at low coverage
($\sim 0.1$ monolayers) in high surface area substrates finds a
diffusivity of about $\SI{0.015}{\cm\squared\per\s}$ at
$\SI{40}{\milli\kelvin}$~\cite{PhysRevLett.85.1468}. These measurements had
5 monolayers of $^4$He below the $^3$He, which is sufficient to be
a superfluid and avoid localizing the $^3$He atoms. For simplicity
we can approximate the capture perimeter as an ellipse with a minor
axis of \SI{1}{\nm} and major axis of \SI{6}{\nm} and determine an
effective isotropic capture cross section~\cite{Crowell1976}.
Assuming electrons are spaced \SI{0.2}{\micro\meter} in Y, and the
channels are spaced \SI{0.2}{\micro\meter} in X (so electrons are
\SI{0.4}{\micro\meter} apart), we use this electron (trap) density, the
capture cross section, and thermal velocity to calculate a capture time
of about \SI{100}{\ns} and a diffusion length of about \SI{1}{\micro\meter}.
Thus, a reasonable density of electrons can rapidly capture the $^3$He,
and the location of where the $^3$He arrived can be determined with a
few micron accuracy. Micron-level resolution is unlikely to be
necessary, and a lower density of channels and electrons should be adequate.
For example, if the electrons are spaced \SI{100}{\micro\meter} apart
in both X and Y, the capture time for a $^3$He atom becomes about
\SI{0.1}{\s}, and the spatial resolution is about \SI{0.4}{\mm}.

\paragraph{$^3$Helium transport}---\,
\AD
{
An important feature of this detector concept is the ability to collect
$^3$He atoms over a large area and bring them to one or a few optimized
quantum sensors.
As discussed earlier, phonons are created in
the bulk of a detector crystal, and they rapidly disperse the energy
throughout the volume
of the material, making their direct detection challenging.
Our concept uses CCDs for $^3$He transport.
In the inset of Fig.~\ref{fig:energy-ccd} we
show a CCD which has been used to demonstrate essentially perfect charge-transfer
efficiency~\cite{PhysRevLett.107.266803} for electrons bound to helium.
In this device the channels run horizontally, while the
gate electrodes can be seen running vertically under the channels.
As the gate voltages are controlled to move the electrons,
they will drag the $^3$He atoms along with
them in moving dimples.
}
{
The CCD device in the inset of Fig.~\ref{fig:energy-ccd} was
made at a standard silicon processing facility, which can fabricate similar
electrode structures over large areas. The assumed 1 kg NaI crystal forms a disk about 2 cm thick and 13 cm in diameter,
and thus the collector must be similar in size. Silicon devices with areas in
that range are practical. Either a single large device could be fabricated, or a number
of smaller ones can be tiled together, as is done for large-area optical CCDs.
}
A variety of experiments have been performed with similar channel structures, including isolating individual and pairs of electrons~\cite{doi:10.1063/1.1900301, 2014}.

\paragraph{$^3$He Detection}---\,
After $^3$He atoms have been evaporated, captured, and collected with the
CCD, it is necessary to detect single atoms. An electron's
spin without a $^3$He trapped in the dimple below it is expected to have
long phase coherence, since the spin-orbit interaction for an electron in
the vacuum is particularly small~\cite{PhysRevA.74.052338}. However, if a
$^3$He atom is trapped by an electron, the nuclear spin will rapidly decohere
the spin of an electron initially prepared in a superposition of up and down spin. 
This decoherence will happen in less than a \SI{1}{\milli\second}, while the spin
coherence of electrons bound to helium is thought to be at least
seconds~\cite{PhysRevA.74.052338}.
This is a quantum nondemolition process and can be repeated as long as the $^3$He
remains trapped, allowing multiple interrogations to ensure reliable detection.
Several approaches to detecting the spin of single electrons bound to helium
are under active investigation, driven by quantum computing applications, and are discussed in the Appendix.
\AD
{
Once the measurement is complete, any
detected $^3$He atoms will be clocked with the CCD to
a region with a large number of electrons tightly bound in circular
($\sim$200 nm diameter) "quantum dots". The $^3$He atoms will be
trapped and gettered by these electrons. Residual $^3$He atoms present
when the detector is initialized will similarly be collected and moved
to the getter region
with the CCD. A 100 nm thick enriched He layer over the
collector area (for the 13 cm diameter target)
can be expected to have $\sim$10$^7$ $^3$He atoms, but it is
quite straightforward to fabricate 10$^8$ or more quantum dots in
an area of $\sim$10 mm$^2$, and each dot can trap multiple $^3$He atoms.
}

As noted above, a cursory evaluation of detector crystals
suggests that NaI may be
appropriate.  The estimate of
2 DM scattering events/minute, with the probability of detecting an event
being about 35\%, compares favorably with the most important
expected background dominated by $^{40}$K decays which will occur about
about once per hour. More detail about the expected detector
backgrounds is discussed in the Appendix. The
generation of scintillation photons as well as
large numbers of phonons in one area in a short
time can be used as a veto for high energy events, such as
radioactive decays and Compton scattering.  The frequency of readout
operation cycles with $^3$He collection,
quantum sensing, and $^3$He gettering will be adjusted
to match event and background rates. 
\AD
{
Readout times of a few ms are
slow, on the typical scale of CCDs and electronics, and thus the heat load
is expected to be small enough to allow operation at 35 mK, or colder. The
heat load could be further reduced with superconducting metallization on the
$^3$He detector structure, though this is probably unnecessary since nearly all the
power will be dissipated in driver circuitry at higher temperature.
}

\paragraph{Summary}---\,
In summary we have presented a new concept for detecting low energy
($\sim$meV) excitations, in particular those which might be generated
in target materials through the interaction with low-mass dark matter.
The approach begins with a DM interaction producing phonons in an ionic crystal, which cause
the quantum evaporation of $^3$He from the surface. The $^3$He is
then caught
on an adjacent surface, where there is an isotopically enriched van
der Waals $^4$He film covering a layer of metallic electrodes and
etched micro-channels holding electrons on the film. We calculate that the
electrons on liquid helium can trap the $^3$He atoms, and they
will drag $^3$He atoms as they are clocked across the helium surface
in a CCD, allowing $^3$He atoms to be collected for detection by
quantum sensors. We suggest that the spin of $^3$He atoms can be coupled
to electron spins for sensitive detection --- to the level of a single
$^3$He atom. Thus the difficult balance of efficient detection of very
rare low-energy events occurring throughout a large volume is
solved in our
approach through the trapping, collection, and quantum sensing
of the $^3$He atoms.
\HL
{
Calculations of dark photon mediated interactions and estimates of
the various background processes show that with a kg-sized ionic
crystal a detected DM event rate of about 40/hr can be achievable,
while high-energy radioactive decays and Compton events will be
about 50 times less frequent. These high-energy events can be
distinguished by the detector, and thus vetoed. Coherent photon
and neutrino scattering will produce low-energy events, similar
to DM, but their estimated rates are over 3 orders of magnitude
less than the DM.
}
Assemblies of dark matter sensors of this design could
operate for long periods with periodic readout of accumulated $^3$He
atoms

All major aspects of this detector concept are based on
established experimental results, or in the cases of single
spin measurement and $^3$He trapping (also
suggested for electron bubbles~\cite{PhysRev.180.259}),
they are being
actively pursued in the context of quantum computer development with electrons on liquid helium~\cite{PhysRevA.74.052338}. Experimental
verification of spin measurement and $^3$He trapping will enable
first-generation detectors and open the door to this path of quantum sensing of phonons for DM detection.

\section*{Acknowledgments}
This work was supported by Quantum Information Science Enabled Discovery
(QuantISED) for High Energy Physics and by the Office of Science of the
U.S. Department of Energy under Contract No. DE-AC02-05CH11231.

\section{Appendices}
\paragraph{Appendix 1. Detector Crystal and Background}---\,
An initial phonon with an energy of a few 10s of meV created in
a dark matter scattering event typically decays through
a sequence of inelastic processes to acoustic phonons with
frequencies of order 1~THz, where thermalization is slowed
in high-quality crystals by the decreasing phonon density
of states.~\cite{wolfe_1998} We will consider light DM detection by
a 1 kg NaI crystal with $^3$He
quantum sensing of the resulting phonons. Other crystals may prove to
be superior, but from a cursory look NaI satisfies several criteria:
(1) it has low energy cut-off ($\sim$20 meV) for phonons generated by DM~\cite{Trickle:2019nya}; (2) it can be purified
to have a low radioactive background;
(3) neither Na nor I have multiple naturally occurring isotopes, thus
eliminating isotopic scattering of the acoustic phonons; and (4) it can
be cleaved, which will reduce the phonon thermalization at surfaces
and may increase the yield of evaporated $^3$He atoms. From calculations
of the cross section for DM interaction within a dark photon
interaction model and a freeze-in model of the DM flux~\cite{PhysRevD.101.055004},
one finds that the rate of DM events is about 2/minute at a DM mass of
about 20 keV in 1 kg of NaI with a minimum energy cut-off of 20 meV.
A 20 meV phonon in the NaI will decay to about
20 acoustic phonons with enough energy to quantum evaporate the $^3$He.
If we assume that the efficiency for an acoustic phonon to desorb a
helium atom is $\sim 5\%$~\cite{PhysRevLett.54.2034,More1996}, and
the probability of that atom
being a $^3$He is about 1/3~\cite{Warren2000}, there
is thus about a 1/60 chance of a single
acoustic phonon being detected through $^3$He evaporation. With each
DM event producing $\sim$20 acoustic phonons, we estimate about one
$^3$He atom will be produced every 1.5 minutes. Improved
preparation of the NaI surfaces or better ionic crystals may
increase the $^3$He evaporation rate.

Backgrounds for this detector are expected to be similar to those seen
by other DM experiments. The DAMA/LIBRA and KIMS
experiments~\cite{damalibra,KIMS_2017}) have established that an
important background source in NaI is residual $^{40}$K. Large NaI crystals
with no more than 20 ppb of potassium impurities~\cite{damalibra} imply a decay rate
of about 1.2/kg/hr. The decay of cosmogenic tritium will also contribute to the
background.  The CDMSlite experiment~\cite{AGNESE_tritium_2019} has found a
tritium production rate of $\sim$75 atoms/kg/day
in a Ge detector.  
Calculations of the tritium production for NaI find a rate of
about 83 atoms/kg/day at sea level.~\cite{tritium_calc2017}
Assuming 60 says at sea level for detector crystal preparation before
installation underground, the cosmogenic tritium will contribute about
one decay every 30 hours.
\HL
{
Each decay of $^{40}$K and tritium  will generate 
many phonons and thus many $^3$He atoms. If the detector is read out
more frequently than these background events, the large signals can be
used as a veto. These high energy events will also produce scintillation
photons which can provide another
avenue for vetoing them. Taken together it is anticipate that there will be about 50 detectable DM events
between $^{40}$K and tritium decays under
the assumptions discussed above. Since these decays can be
vetoed based on their deposited energy, they will contribute
to detector dead time, but will not otherwise interfere
with the DM signal.
}
\HL
{
Compton scattering of MeV-scale photons will
deposit high energies in the detector crystal,
which can be vetoed as described above,
but Robinson has pointed out that coherent photon
scattering can deposit much smaller energies, of the same
order as DM events.~\cite{Robinson:2017}  He has
calculated an integrated scattering rate of $\sim$0.34
events/kg/day for recoil energies below 1 eV in Ge,
assuming a well-constructed passive radiation
shield and neglecting both coherence between atoms
and phonon quantization in the crystal. Again, higher
energy events can be vetoed. The iodine in the NaI crystal will
dominate the coherent photon scattering, having a 3.8x larger atomic cross section than Ge through the relevant
energy range ~\cite{Chatterjee:1998}. Under similar assumptions
we estimate that coherent photon scattering will produce
$\sim$0.6 recoils/kg/day. Being of similar energy as the DM events,
it is not possible to veto these recoils, but their rate is about
3 orders of magnitude lower than the calculated DM rate
(2/kg/min. in NaI). Coherent neutrino scattering will
similarly generate recoils which cannot be vetoed based on
deposited energy. However, for recoil energies below $\sim$1 eV,
the coherent photon scattering rates as estimated by
Robinson exceed the expected
coherent neutrino rates.~\cite{Robinson:2017,Billard:2014} Thus, while the photons and neutrinos will add a small offset to the
DM signal, this background is expected to be smaller by several
orders of magnitude.
}

\AD
{
The collector
device can also introduce backgrounds from radioactive decay.
It appears best to avoid
Al metallization in the collector chip, since cosmogenic $^{26}$Al could
add a considerable background. A copper process will
avoid this issue. If the collector
is made as a standard silicon device, it will also introduce signals from the decay of
cosmogenic $^{32}$Si, as has been seen in other DM experiments. The DAMIC
experiment~\cite{DAMIC_32Si_2015}
has quantified the radioactivity of $^{32}$Si, and finds it contributes
about 80 decays/kg/day.  A typical Si wafer with the 13 cm diameter discussed above
weighs about 30 gm, and thus the $^{32}$Si can be expected to cause about 3 events/day.
Again, these are high energy events which can be vetoed.
Being considerably lighter than the target crystal, the tritium background from
the collector is not expected to be a major contributor, with a rate comparable to $^{32}$Si.
}

\AD
{
Other background sources for this class of quantum evaporation detectors have been
identified and modeled as part of the HeRALD experiment,~\cite{PhysRevD.100.092007}
including the layers of shielding required. Background excitation of the helium
is suppressed by its large bandgap for electronic excitations. NaI has a smaller
gap, of about 5.8 eV, but  most of that analysis carries over to
this case.  The gap still protects against low-energy processes. If an
event does excite an electron across the gap, a large number of phonons will be
produced when the electron-hole pair recombine or are trapped,
and again these events can be identified. There is evidence in some other
ultra-sensitive DM detection devices that stresses built up in materials can
slowly relax by emitting phonons. In the detectors discussed
here no thin films are deposited on the target crystals. Such films can undergo thermal expansion mismatch stresses,
though low-stress mounting will still be important.
}

\paragraph{Appendix 2. $^3$He Detection}---\,
Multiple approaches are being taken by different groups for measuring
spins on helium.  Detection of single nuclear spins in other systems
has been accomplished with
quantum sensors in recent years.  For example, nitrogen-vacancy centers
have been used to sense the presence of nearby $^{29}$Si
atoms~\cite{mueller2014}. However, it is not clear whether direct nuclear
spin detection can be adapted to the situation of a $^3$He atom on
$^4$He, since the direct sensing of nuclear spins has relied on extremely
close and stable positioning of the nucleus and the sensor.
Converting to an electron spin, with its much larger magnetic
moment, appears easier as discussed in the main text. Detection of
single electron spins has been demonstrated in a range of quantum
sensor and qubit platforms, from quantum dots to color
centers~\cite{RevModPhys.89.035002}.  It has
been shown that the electron motion can be coupled to a superconducting
resonator with a coupling constant of $\sim \SI{5}{\MHz}$~\cite{koolstra2019}.
However, these first experiments were limited by decoherence of the
motional states, apparently due to vibrations exciting fluctuations
in the helium surface. Recently, strong coupling of the electron
motion to a superconducting micro-resonator while bound to solid
neon has been demonstrated~\cite{dafei2021}. Isolating the helium
from the vibrations is being investigated in several labs and a high
degree of vibration isolation will be central to the integration of
our detector concept. With the motion strongly coupled to the resonator,
an inhomogeneous magnetic field can provide the spin interaction, as
has been demonstrated for electrons in silicon quantum
dots~\cite{mi2018, samkharadze2018}.  In an alternative configuration, one
could utilize a pair of electrons initialized to a spin singlet in a
nanofabricated quantum dot, separating the two electrons, trapping the
$^3$He under one to shift its phase, and then bringing the electrons
back together to determine whether they are still a singlet.
Decoherence from (single) $^3$He atoms will drive them from the singlet
to the triplet with $m = 0$.  A third approach would be to use a color
center, like a NV$^-$ or SiV$^0$ in diamond to sense the electron
spin (much less demanding than sensing a nuclear spin)
~\cite{jakeT2008}.
Direct ESR techniques may also be possible, where sensitivity
to a single electron's spin has recently been
demonstrated.~\cite{wang-bertet:2023}
The signal could be enhanced by using one
$^3$He atom to sequentially decohere multiple electrons, since the atom
is preserved in the process (its spin need not be preserved).

\AD
{
Here we have concentrated on using the $^3$He nuclear spin for quantum
sensing, but there may be other ways to utilize the unique signatures of $^3$He.
For example, the CCDs could be arranged to transport all of the $^3$He atoms
to one place, where they are ejected from the surface with a heat
pulse. With the atoms all emerging in one place, an ionization process
like that described by Maris, {\em et al.}~\cite{PhysRevLett.119.181303},
but with isotope-selective
(perhaps optical) excitation, could be employed. Alternatively, ultra-sensitive
mass spectrometry or other sensing technique might be enabled with the
localized He source. 
}
\paragraph{Appendix 3. Alternative Adsorbates}---\,
We have concentrated on the evaporation of $^3$He from the surface
of liquid $^4$He, since it has the lowest surface binding energy
($\sim \SI{5}{\K}$). However, many other atomic and molecular species
as well as electrons can be bound to a liquid He surface, and their
evaporation may prove useful as phonon detectors. An isolated electron
binds with an energy of $\sim \SI{0.6}{\meV}$~\cite{RevModPhys.46.451},
but a high electron density is necessary if the ejection of an
electron is to have a high probability. However, large holding fields
are then required to hold the electrons on the surface, and electron
emission is limited by electron-electron interactions~\cite{Kono1982}.
Alkali metals were predicted to bind to helium with energies of
$10 \sim \SI{20}{\K}$~\cite{Ancilotto1995}, and experimentally found
to bind to the surface of He nanodroplets~\cite{Stienkemeier1996}. Being
uncharged they do not require holding fields, but at high densities they
form dimers and clusters. It has also been reported that other species,
such as HD, can be desorbed from alkali halides with a single
phonon~\cite{HDonLiF1987}.  Such species may be useful as detectors for
particular energy ranges of proposed dark matter candidates and interactions.
Being much more polarizable than He, it may also be possible to tune their
desorption energy with an applied electric field.

\newpage

\bibliographystyle{apsrev4-1} 
\bibliography{references} 

\end{document}